\documentclass[10pt,conference,letterpaper,twoside,twocolumn]{IEEEtran}

\usepackage{cite}
\usepackage{graphicx,color,epsfig,rotating}
\usepackage{amsfonts,amsmath,amssymb}
\usepackage{algorithm,algorithmic}
\usepackage{subfigure}
\long\def\comment#1{}

\newfont{\bbb}{msbm10 scaled 700}

\newfont{\bb}{msbm10 scaled 1100}

\newcommand{\RR}{\mbox{\bb R}}

\newcommand{\ZZ}{\mbox{\bb Z}}

\newcommand{\EE}{\mbox{\bb E}}

\newcommand{\av}{{\bf a}}

\newcommand{\cv}{{\bf c}}

\newcommand{\hv}{{\bf h}}

\newcommand{\uv}{{\bf u}}
\newcommand{\wv}{{\bf w}}

\newcommand{\xv}{{\bf x}}

\newcommand{\zv}{{\bf z}}
\newcommand{\zerov}{{\bf 0}}

\newcommand{\Bm}{{\bf B}}

\newcommand{\Gm}{{\bf G}}

\newcommand{\Id}{{\bf I}}

\newcommand{\Lm}{{\bf L}}

\newcommand{\Cc}{{\cal C}}

\newcommand{\Nc}{{\cal N}}

\newcommand{\Sc}{{\cal S}}

\newcommand{\Vc}{{\cal V}}

\newcommand{\lambdav}{\hbox{\boldmath$\lambda$}}

\newcommand{\Lambdam}{\hbox{\boldmath$\Lambda$}}

\newcommand{\Sigmam}{\hbox{\boldmath$\Sigma$}}

\newcommand{\SNR}{{\sf SNR}}

\newcommand{\transp}{{\sf T}}

\newtheorem{theorem}{Theorem}
\newtheorem{lemma}{Lemma}

\begin{document}

\title{Quantized Compute and Forward: \\ A Low-Complexity Architecture for Distributed Antenna Systems}

\author{
\IEEEauthorblockN{Song-Nam Hong}
\IEEEauthorblockA{
University of Southern California \\
Los Angeles, CA 90089\\
songnamh@usc.edu}
\and
\IEEEauthorblockN{Giuseppe Caire}
\IEEEauthorblockA{
University of Southern California \\
Los Angeles, CA 90089 \\
caire@usc.edu}
\thanks{This research was supported in part by NSF Grant CCF-CIF 0917343}
}

\maketitle

\begin{abstract}
We consider a low-complexity version of the \emph{Compute
and Forward} scheme that involves only scaling, offset
(dithering removal) and scalar quantization at the relays. The
proposed scheme is suited for the uplink of a distributed antenna
system where the antenna elements must be very simple and are
connected to a joint processor via orthogonal perfect links of
given rate $R_{0}$. We consider the design of \emph{non-binary} LDPC codes
naturally matched to the proposed scheme. Each antenna element
performs individual (decentralized) Belief Propagation decoding
of its own quantized signal, and sends a linear combination of
the users'  information messages via the noiseless link to the
joint processor, which retrieves the users' messages by Gaussian
elimination. The complexity of this scheme is linear in the coding
block length and polynomial in the system size (number of relays).
\end{abstract}


\section{Introduction}\label{sec:intro}

Powerful multi-media capable devices such as smartphones and tablets demand higher and higher
data rates. Without increasing the system bandwidth (a scarcely available commodity) and the geometry
of the cells, this demand is accomplished by more and more sophisticated signal processing.
In turn, this trend makes {\em energy efficiency} one of the most pressing problems in modern wireless cellular networks.
It is known that the power consumption of conventional tower-mounted macro-BS
in today's high-speed data oriented systems (e.g., 3G HSDPA/HSUPA  and 4G LTE \cite{Molisch})
contributes for a large fraction of the operational costs of wireless cellular operators \cite{Etoh}.
Furthermore, the environmental impact of cellular networks is striking: for example,
the carbon footprint of downloading a 1.5GB video file (typical iTunes movie size) from a 3G/HSDPA tower
is $\sim$ 22 kg-CO$_2$e \cite{EARTH}, comparable with driving an SUV for 40 miles and significantly larger than downloading the same amount of bits
from a wired link ($\sim$ 0.75 22 kg-CO$_2$e).

An alternative approach to the conventional BS architecture consists of a ``cloud'' BS, where a large number of
simple single-antenna elements are distributed over a wide area and connected to a central processor
via wired backbone links. The distributed antenna elements perform demodulation, Analog to Digital Conversion (ADC), and possibly
some decentralized decoding operation, and ship their processed observations
to the central processor, which performs some form of joint decoding.  In this way, any {\em User Terminal} UT finds a BS antenna element at small distance with high
probability, thus allowing for a very dense spatial reuse.
In terms of power consumption, many small terminals are significantly more power efficient than
a single large one (e.g., no need for power-consuming cooling subsystems). Furthermore, additional power savings can be
obtained by switching off the antenna elements not in the vicinity of ``active'' users.
Since in practice cellular users have very large ``off'' duty cycles, a correspondingly large fraction of antenna elements  is switched
off at any given time.

An uplink distributed antenna system with $L$ users and $L$ single-antenna terminals, connected
to a central  processor via wired links of fixed rate $R_0$  forms a three layer relay network with $L$ sources, $L$
relays and one destination.  This scenario was investigated in \cite{Sanderovich1}, where Decode-and-Forward (DF)
and Compress-and-Forward (CF) strategies were analyzed in closed form for the so-called {\em Wyner model}
(see Section \ref{wynermodel}). The DF strategy makes the central processor extremely simple but it is not compliant with the basic goal of having
very low-complexity distributed antenna elements. The CF strategy of \cite{Sanderovich1} is in fact a special case of the so-called ``quantize-remap and forward''
scheme of \cite{Avestimehr}, also referred to as ``noisy network coding'' in \cite{Lim}.
In this case, the relays (i.e., the distributed antenna elements) just quantize the received signal
and forward the quantization samples.\footnote{The information-theoretic vector quantization
of \cite{Avestimehr}, \cite{Lim} can be replaced by scalar quantization with a fixed-gap performance degradation \cite{Ozgur}.}
However, the central processor is required to jointly decoding all users from the quantized observations.

An alternative approach consists of {\em Compute and Forward} (CoF),
recently developed in \cite{Bobak_CF}, \cite{Feng}, and applied
to the Wyner model in \cite{Sanderovich2}. CoF seeks a tradeoff between ``quenching'' the noise at the relays (as DF) and
forwarding ``noisy'' observations without making local decision (as CF).
In this scheme, the users encode their information messages using the {\em same} lattice code, and each relay decodes a linear
combination with integer coefficients of the  lattice codewords. These linear combinations are mapped onto linear combination of the messages defined
on a suitable finite field and forwarded to the central processor. If no error at the relays occurs and the overall $L \times L$ linear system has rank $L$,
the central processor can decode the user messages  in polynomial time by Gaussian elimination, as in standard linear network coding \cite{Koetter}.

In this paper we focus on the uplink of a distributed antenna system and consider a CoF architecture
suited for low-complexity low-power system implementation.  The proposed scheme is motivated by the observation that the main bottleneck
of a digital receiver is the {\em Analog to Digital Conversion} (ADC), which is costly, power-hungry and
does not scale with Moore's law.  Rather the number of bit per second produced by an ADC is roughly a constant that depends
on the power consumption \cite{Walden,Singh}.  Therefore, it makes sense to consider the ADC transformation as part of the channel
transformation, which becomes discrete-output in nature.   We propose a {\em Quantized CoF} (QCoF) scheme based on one dimensional \emph{lattice modulation} and linear codes
over $\ZZ_{p}$ ($p$ being a prime number).  Each relay recovers a noisy linear combination of the user codewords,
where the noise is discrete and additive over $\ZZ_p$. Therefore, the resulting {\em computation rate} coincides with the
capacity of a single-user discrete additive noise channel, achieved by linear codes over $\ZZ_{p}$ \cite{Dobrushin}.
For large $p$, QCoF achieves the computation rate of unquantized CoF within the shaping loss, about $0.25$ bits per real symbol.

We also develop practical Low-Density Parity-Check (LDPC) code constructions  for the additive noise channel over $\ZZ_p$,
inspired by the work of \cite{Bennatan}. We show results for an ensemble of regular and irregular Repeat-Accumulate (RA) protograph-based
codes over $\ZZ_p$, the performance of which is evaluated via the EXIT-chart method.
More refined code design (e.g., based on the ``ARA'' code structure \cite{Abbasfar}) is expected to provide further improvements for
a wider range of system parameters.

We compare the spectral efficiency of QCoF with the benchmarks provided by DF, CF, and CoF, over the standard Wyner model. The proposed scheme shows competitive performance with respect to much more involved schemes, which rely on infinite ADC resolution. Also, we show that unequal power allocation can be used to reduce the ``non-integer" error of QCoF and further improve its performance.

\section{Quantized compute and forward}\label{sec:SCF}

\subsection{Definitions}

Let $\ZZ_{p} = \ZZ \mod p\ZZ$ denote the finite field of size $p$, with $p$ a prime number, $\oplus$ denote addition over $\ZZ_p$,
and $g : \ZZ_p \rightarrow \RR$ be a function that maps the elements of $\ZZ_p$ into the points $\{0,1,...,p-1\} \subset \RR$.

For a lattice $\Lambda$, we define the lattice quantizer $Q_{\Lambda}(\xv) = {\rm argmin}_{\lambdav \in \Lambda} \{  \| \xv - \lambdav\| \}$,
the Voronoi region $\Vc = \{ \xv \in \RR^n : Q_\Lambda(\xv) = \zerov\}$  and
$[\xv ] \mod \Lambda =  \xv - Q_{\Lambda}(\xv)$.
Let $p$ be a prime integer and $\kappa \in \RR_+$. We consider the two nested one-dimensional lattices
\begin{eqnarray}
\Lambda_s  & = & \{x = \kappa p z: z \in \ZZ\} \nonumber \\
\Lambda_c  & = & \{x = \kappa z: z \in \ZZ\}.
\end{eqnarray}
and define the {\em constellation set} $\Sc \triangleq \Lambda_c \cap \Vc_s$, where $\Vc_s$ is the {\em Voronoi region} of $\Lambda_s$, i.e., the interval
$[-\kappa p/2, \kappa p/2)$.  The {\em modulation mapping} $M : \ZZ_p \rightarrow \Sc$ is defined by
$v = M(u) \triangleq [\kappa g(u)] \mod \Lambda_s$. The inverse function $M^{-1}(\cdot)$ is referred to as the
{\em demodulation mapping}, and it is given by  $u = M^{-1}(v) \triangleq g^{-1}( [ v/\kappa ]  \mod p \ZZ)$ with $v \in \Sc$.

Let $\Gm \in \ZZ_p^{n \times k}$ be a matrix of rank $k$.
The linear code $\Cc$ over $\ZZ_p$ with block length $n$, dimension $k$ and rate $R = \frac{k}{n} \log(p)$ (in bit/symbol) generated by $\Gm$ is the set of codewords
$\Cc = \{\cv  = \Gm \wv : \wv \in \ZZ_p^k\}$.

\subsection{Modulation and coding for QCoF}

Consider the (real-valued) $L$-user Gaussian multiple access channel with inputs $\{x_{\ell,i} : i = 1, \ldots, n\}$ for
$\ell = 1,\ldots, L$, output $\{y_i : i = 1,\ldots, n\}$ and
coefficients  $\hv = (h_1,\ldots, h_L)^\transp \in \RR^L$, defined by
\begin{eqnarray} \label{realMAC}
y_{i}  = \sum_{\ell=1}^{L} h_{\ell} x_{\ell,i} + z_i, \;\;\; i = 1,\ldots, n
\end{eqnarray}
where the $z_i$'s are i.i.d. $\sim \Nc(0,1)$.
For this channel, we consider a modification of the CoF scheme of \cite{Bobak_CF} that includes scalar quantization
at the receiver as part of the channel.
In the proposed scheme, all users encode their information messages $\{\wv_\ell \in \ZZ_p^k : \ell = 1, \ldots, L\}$ using the same
linear code $\Cc$ over $\ZZ_p$, and produce their channel inputs according to
\begin{equation} \label{input}
x_{\ell,i} = [ M(c_{\ell,i}) + d_{\ell,i}] \mod \Lambda_s
\end{equation}
for $i=1,...,n$,  where $c_{\ell,i}$ is the $i$-th component of codeword $\cv_\ell = \Gm\wv_\ell$ and
the $d_{\ell,i}$'s are i.i.d. dithering symbols  $\sim \mbox{Uniform}(\Vc_s)$, known at the receiver.  The channel inputs $x_{\ell,i}$ are uniformly
distributed over $\Vc_s$ and have second moment $SNR \triangleq \EE[|x_{\ell,i}|^{2}] = \kappa^2 p^2 / 12$.

The receiver's goal is to recover a linear combination $\cv = \bigoplus q_{\ell} \cv_{\ell}$ of the transmitted users' codewords, for some
coefficients $q_\ell \in \ZZ_p$. For this purpose, the receiver selects the \emph{integer coefficients vector} $\av= (a_{1},...,a_{L})^{\transp} \in \ZZ^{L}$
and produces the sequence of quantized observations
\begin{equation}\label{eq:demo}
u_{i} =   M^{-1} \left ( \left [Q_{\Lambda_c} \left (\alpha y_{i} - \sum_{\ell=1}^{L} a_{\ell} d_{\ell,i} \right ) \right ] \mod \Lambda_s \right ),
\end{equation}
for $i=1, \ldots ,n$.
Letting $\uv = (u_1,\ldots, u_n)^\transp$, and using Lemma \ref{lem:transform} at the end of this section,
we have that  the concatenation of (\ref{input}), (\ref{realMAC}) and (\ref{eq:demo}) is equivalent to
\begin{equation}\label{eq:model}
\uv = \Big( \bigoplus_{\ell=1}^{L} q_{\ell} \cv_{\ell} \Big) \oplus \tilde{\zv},
\end{equation}
with $q_{\ell} = g^{-1}([a_{\ell}] \mod p\ZZ)$,
and where the discrete additive noise vector $\tilde{\zv}$ has statistics given in Section \ref{sec:main-result}.

\begin{figure}
\centerline{\includegraphics[width=9cm]{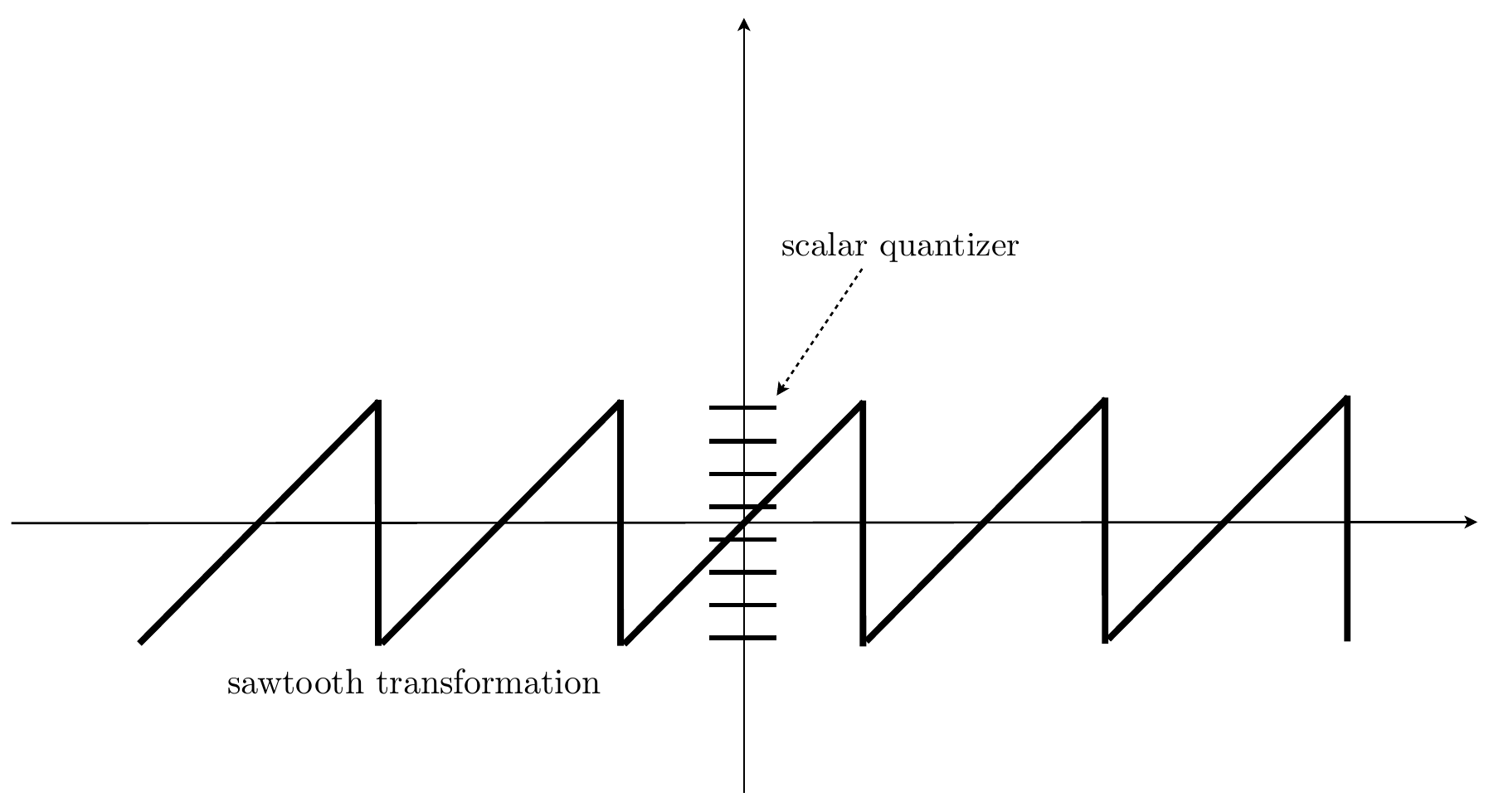}}
\caption{Implementation of $Q_{\Lambda_c}(\cdot)$ followed by modulo $\Lambda_s$ using an analog transformation
and a finite levels scalar quantizer.}
\label{receiver-structure}
\end{figure}

Notice that $\uv$ is obtained by componentwise analog operations (scaling and translation) and scalar quantization.
in fact, the scalar quantization by $\Lambda_c$ followed by the modulo $\Lambda_s$ operation
can be obtained by concatenating a sawtooth memoryless transformation with a standard $p$-points scalar quantizer,
as shown in Fig.~\ref{receiver-structure}. The components $u_i$ of $\uv$ can be sent to the central processor at the rate of $\log(p)$ bit/symbol.
Alternatively, the linear combination $\cv$ can be locally decoded, and the corresponding linear combination of the user messages,
$\wv = \bigoplus q_{\ell} \wv_{\ell}$ can be sent to the central processor at rate $k/n$ bits/symbol.

Letting $D: \ZZ_{p}^{n} \rightarrow \Cc$ denote a decoder for $\Cc$,
we define the average error probability as $P_e(\hv,\av) = P(D(\uv) \neq \cv)$, for fixed
coefficients $\hv, \av$, averaged over the messages,  the channel noise and the dithering signals.
A {\em computation rate} $R(\hv,\av)$ for the QCoF scheme described above is achievable if there exist
a sequence of $(n,k)$ codes $\Cc$ such that $\liminf_{n \rightarrow \infty} \frac{k}{n} \log(p) \geq R(\hv,\av)$
and $\lim_{n \rightarrow \infty} P_e(\hv,\av) = 0$.

\begin{lemma}\label{lem:transform}
For $u_{\ell} \in \ZZ_{p}$, let $u =\bigoplus_{\ell=1}^{L} q_{\ell} u_{\ell}$
for some coefficients $q_{\ell} \in \ZZ_{p}$.
Also, set $v_{\ell} = M(u_{\ell})$ and $v = \Big[ \sum_{\ell=1}^{L} a_{\ell} v_{\ell} \Big] \mod \Lambda_s$
for some $a_{\ell} \in \ZZ$ such that $q_{\ell} = g^{-1}([a_{\ell}] \mod p\ZZ)$. Then, we have $u = M^{-1}(v)$.
\hfill $\square$
\end{lemma}

\section{Achievable computation rate}\label{sec:main-result}

First we examine the marginal statistics of the discrete noise $\tilde{\zv}$ in (\ref{eq:model}).
Rewriting (\ref{eq:demo}) by letting $v_{\ell,i} = M(c_{\ell,i})$ and by omitting the subscript $i$ for brevity, we have
\begin{align}
& v =  \Big[Q_{\Lambda_c} \Big( \alpha y - \sum_{\ell=1}^{L} a_{\ell} d_{\ell} \Big)\Big] \mod \Lambda_s \nonumber \\
& = \Big[Q_{\Lambda_c} \Big( \alpha \sum_{\ell=1}^L h_\ell x_\ell + \alpha z  - \sum_{\ell=1}^{L} a_{\ell} d_{\ell} \Big)\Big] \mod \Lambda_s \nonumber \\
& = \Big[Q_{\Lambda_c} \Big( \sum_{\ell=1}^L a_\ell v_\ell  + \sum_{\ell=1}^L (\alpha h_\ell  - a_\ell) x_\ell + \alpha z  \Big)\Big] \mod \Lambda_s \nonumber \\
& = \Big[Q_{\Lambda_c} \Big( \sum_{\ell=1}^L a_\ell v_\ell  + \sum_{\ell=1}^L e_\ell + \alpha z  \Big)\Big] \mod \Lambda_s
\end{align}
where we let $e_{\ell} = (\alpha h_{\ell} - a_{\ell})x_\ell$ and where we used the fact that, by (\ref{input}),
$\sum_{\ell=1}^L a_\ell (v_\ell + d_\ell)$ and $\sum_{\ell=1}^L a_\ell x_\ell$ differ by some point of $\Lambda_s$ and,
for any $y \in \RR$ and $\lambda \in \Lambda_s$, we have $Q_{\Lambda_c}(y + \lambda) = Q_{\Lambda_c}(y)$.

By the well-known Crypto-Lemma \cite{Forney},
the non-integer error term $e_{\ell}$ is statistically independent of $v_{\ell}$ and is uniformly distributed in
$[-\sqrt{3\SNR}(\alpha h_{\ell}- a_{\ell}),\sqrt{3\SNR}(\alpha h_{\ell} - a_{\ell}))$.  The overall error and noise term is
$\varepsilon \triangleq \sum_{\ell=1}^{L} e_{\ell} + \alpha z$, with mean zero and variance
\begin{eqnarray} \label{eq:noiseVar}
\sigma_{\varepsilon}^2 & = & \SNR  \left \| \alpha\hv - \av \right \|^2 + \alpha^2.
\end{eqnarray}
The  components of the \emph{effective noise} $\tilde{\zv}$ in (\ref{eq:model}) are distributed as
\begin{equation}
\tilde{z} =  M^{-1}\left (\left [ Q_{\Lambda_c}(\varepsilon) \right ] \mod \Lambda_s \right )
\end{equation} and have pmf that can be calculated numerically, and it is well approximated by assuming $\varepsilon \sim \Nc(0,\sigma_{\varepsilon}^2)$.

Since decoding the linear combination $\sum_{\ell=1}^L q_\ell \cv_\ell$
from the noisy discrete observation (\ref{eq:model}) is equivalent to decoding the linear code $\Cc$
over the additive noise discrete channel $\uv = \cv \oplus \tilde{\zv}$, we have:

\begin{theorem}\label{thm:COMP}
For given $\hv, \av, \alpha$ and modulation order $p$, the largest achievable computation rate
of QCoF is equal to the capacity of the discrete additive noise channel
$\tilde{y} = \tilde{x} \oplus \tilde{z}$, given by  $R(\hv,\av,\alpha) = \log{p} - H(\tilde{z})$.
\hfill $\square$
\end{theorem}

The QCoF computation rate can be maximized  by  minimizing the entropy $H(\tilde{z})$ with respect to $\av \in \ZZ^L, \alpha$ (as a function of $\hv$, $\SNR$ and $p$).
This is generally a difficult problem.  Instead, we resort to the suboptimal (but much simpler) problem of minimizing the variance
$\sigma_\varepsilon^2$ in (\ref{eq:noiseVar}).
First, it is immediate to see that the optimal $\alpha$ is given by \cite{Bobak_CF}
\[ \alpha^{*} = \frac{\SNR\hv^{\transp} \av}{1+\SNR \|\hv \|^2}. \]
Replacing $\alpha^*$ into (\ref{eq:noiseVar}), we obtain
\begin{eqnarray} \label{quadratic}
\sigma_{\varepsilon}^2 & = & \SNR \left (\|\av \|^2 - \frac{\SNR |\hv^{\transp}\av |^2}{1+\SNR \|\hv \|^2} \right ) \nonumber \\
& = & \av ^\transp \left ( \SNR^{-1} \Id  +  \hv \hv^\transp \right )^{-1} \av
\end{eqnarray}
The quadratic form in (\ref{quadratic}) is positive definite for any $\SNR < \infty$, since
the matrix $(\SNR^{-1} \Id  +  \hv \hv^\transp)^{-1}$ has eigenvalues
\[ \lambda_1 = \frac{\SNR}{1 + \SNR \|\hv \|^2}, \;\;\; \lambda_{2} = \cdots = \lambda_{L}=\SNR. \]
By Cholesky decomposition, there exists a lower triangular matrix $\Lm$ such that
$\sigma^2_\varepsilon = \left \| \Lm^\transp \av \right \|^2$. It follows that the problem of minimizing
$\sigma^2_\varepsilon$ over $\av \in \ZZ^L$ is equivalent to finding the ``shortest lattice point'' of the $L$-dimensional lattice
generated by $\Lm^\transp$. This can be efficiently obtained using the LLL algorithm \cite{LLL},
possibly followed by Phost or Schnorr-Euchner enumeration (see \cite{Damen}) of the non-zero lattice points in a sphere centered at the origin,
with radius equal to the shortest vector found by LLL.

\section{Code construction for QCoF} \label{sec:LDPC}

Given the equivalence of the QCoF computation rate with the capacity of the additive noise channel over $\ZZ_p$ with noise $\tilde{z}$,
stated in Theorem \ref{thm:COMP}, our goal is to design low-complexity capacity-approaching linear codes for this channel.
To this purpose, we consider ensembles of protograph-based Low Density Parity-Check  (LDPC) codes \cite{Thorpe}
over $\ZZ_{p}$. In particular, we consider a random coding ensemble where the non-zero elements in the parity-check matrix are chosen
randomly and uniformly chosen from $\ZZ_{p}^*$ (non-zero elements of $\ZZ_p$).
In order to optimize and evaluate the asymptotic (large $n$) performance of these codes
we used  the Gaussian approximation to density evolution known as ``EXIT chart'', as given in \cite{Bennatan}.
Messages generated by the Belief Propagation decoder, in the form of log-likelihood ratios, are modeled as $(p-1)$-dimensional correlated
Gaussian vectors $\Lambdam$ with mean $\nu/2$ and covariance matrix $\Sigmam$ with elements $[\Sigmam]_{i,j} = \nu$
for $i=j$ and $[\Sigmam]_{i,j} = \nu/2$ for $i\neq j$.
Letting $V$ the code variable corresponding to the edge message $\Lambdam$,
we define the mutual information function
\begin{equation}\label{eq:J_fun}
J(\nu) \triangleq I(V;\Lambdam)  =  1-  \EE\Big[\log_{p} \Big( 1 + \sum_{i=1}^{p-1}e^{-\lambda_{i}} \Big)\Big].
\end{equation}
The EXIT chart analysis tracks the evolution of the mutual information $I(V;\Lambdam)$ along the protograph edges \cite{Park}.
A protograph is a bipartite multigraph with $N_c$ ``check'' nodes and $N_v$ ``variable'' nodes, described by a \emph{base matrix}
$\Bm = [b_{i,j}]$ whose element $b_{i,j}$ indicate the multiplicity of the edges connecting the variable node $v_{j}$ to the check node $c_{i}$.
For instance, the protograph corresponding to the Repeat-Accumulate (RA) code ensemble \cite{Jin} of rate $1/2$ has base
matrix $\Bm = [4,2]$.

\subsection{Numerical results} \label{subsec:NR2}

\begin{figure}
\centerline{\includegraphics[width=9cm]{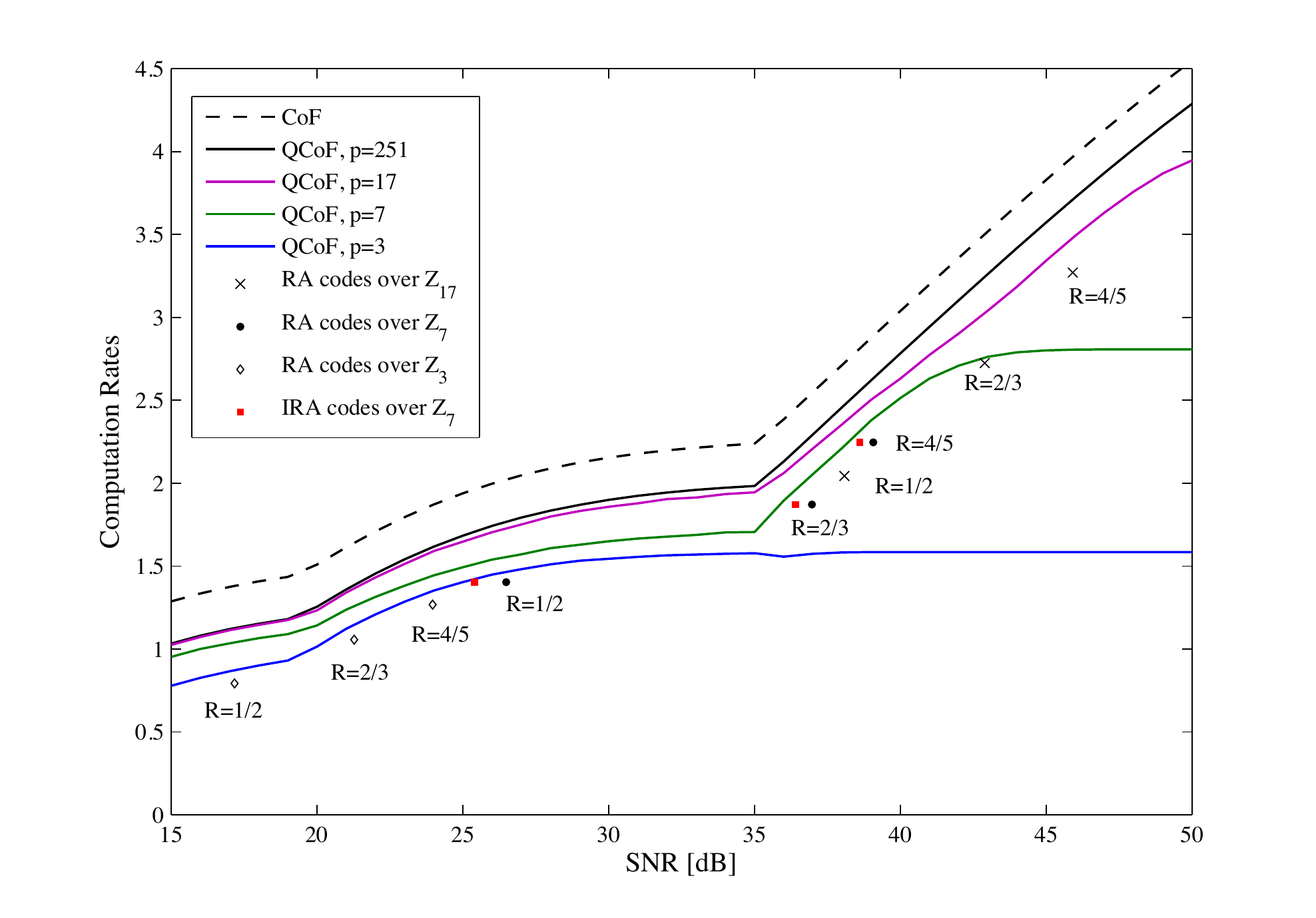}}
\caption{Computation rates for three-user Gaussian MAC with coefficients $\hv = [1,0.75,-\sqrt{2}]$ and ADC level, $p=3$, $p=7$, $p=17$, or $251$.}
\label{simulation_IRA}
\end{figure}

As an example, Fig.~\ref{simulation_IRA} shows the QCoF computation rate as a function of SNR for a $L = 3$ users case. For large $p=251$, QCoF approaches
computation rate of CoF within the shaping loss of $\approx 0.25$ bits/symbol.  For lower $\SNR$,
QCoF with smaller values of $p$ yields satisfactory performance with lower complexity.
We also show the computation rates achieved by a family of RA codes over $\ZZ_{p}$ with code rates $(1/2, 2/3, 4/5)$ for different $p$ values, which performs quite well as shown in Fig.~\ref{simulation_IRA}.
As in the binary case, we can expect that carefully optimized code ensembles can closely approach the information theoretic computation rate
at all SNRs. For example, we designed a simple protograph-based Irregular Repeat-Accumulate (IRA)
code over $\ZZ_{7}$ with rate $1/2$ defined by the base matrix
\begin{equation}
\Bm = \left[
        \begin{array}{cccccccc}
          1 & 0 & 1 & 1 & 1 & 1 & 0 & 0 \\
          0 & 1 & 1 & 1 & 0 & 1 & 1 & 0 \\
          1 & 1 & 0 & 1 & 0 & 0 & 1 & 1 \\
          0 & 0 & 1 & 2 & 1 & 0 & 0 & 1 \\
        \end{array}
      \right].
\end{equation}
Higher coding rates $2/3$ and $4/5$ were obtained by the check node merging technique in \cite{Hong}.
Even this simple IRA code design shows noticeable improvement with respect to its RA counterpart, and almost achieves the theoretical limit at rates $2/3$
and $4/5$. Charts like Fig.~\ref{simulation_IRA} can be used as a guideline to select the modulation order $p$ and coding rate. For instance, at $\SNR =25$dB it is reasonable to choose $p=7$ and an IRA code with rate $1/2$.

\section{Distributed antenna system}\label{wynermodel}

We consider a distributed antenna system with $L$ users and $L$ single-antenna terminals connected to a central processor via wired links of fixed rate $R_{0}$,
as introduced in \cite{Sanderovich1}.
Also, we assume that each $\ell$-th relay is equipped with a $p$-level ADC, as shown in Fig. \ref{receiver-structure}, implementing (\ref{eq:demo}),
followed by a decoder for the linear code $\Cc$, producing the estimated linear combination $\hat{\cv}_\ell$ of the user codewords.
For simplicity and for the sake of comparison with \cite{Sanderovich2}, we consider the symmetric Wyner model for which the
the received signal at the $\ell$-th relay is
\begin{equation}
y_{\ell,i} =  x_{\ell,i} + \gamma( x_{[\ell-1]_{L},i} + x_{[\ell+1]_{L},i}) + z_{\ell,i}
\end{equation}
where $[\ell]_{L} = \ell \mod L$ and $\gamma \in [0,1]$.
The achievable rate of QCoF is
\begin{equation}\label{eq:BQCoF}
R  = \min\left \{\max_{\av, \alpha} R(\hv, \av, \alpha) , R_{0} \right \}
\end{equation}
where $\hv = (\gamma, 1, \gamma)^\transp$ and $R(\hv, \av, \alpha)$ is given by Theorem \ref{thm:COMP}.
The central processor receives via the noiseless links the $L$ linear combinations of the user messages
and,  provided that no decoding error occurred at the relays and that the overall $L \times L$ linear system matrix has rank $L$, it solves for the user
using Gaussian elimination. In this example, thanks to the banded structure
of the channel matrix, the  system matrix is guaranteed to have rank $L$.
Using LDPCs with iterative Belief Propagation decoding, the overall complexity of the receiver scales linearly with the coding
block length $n$ (complexity $O(1)$ per decoded information bit),  and polynomially with the system size $L$.
For the structured banded channel matrix of this example, the complexity per decoded information bit is also $O(1)$
with respect to $L$.

\begin{figure}
\centerline{\includegraphics[width=10cm]{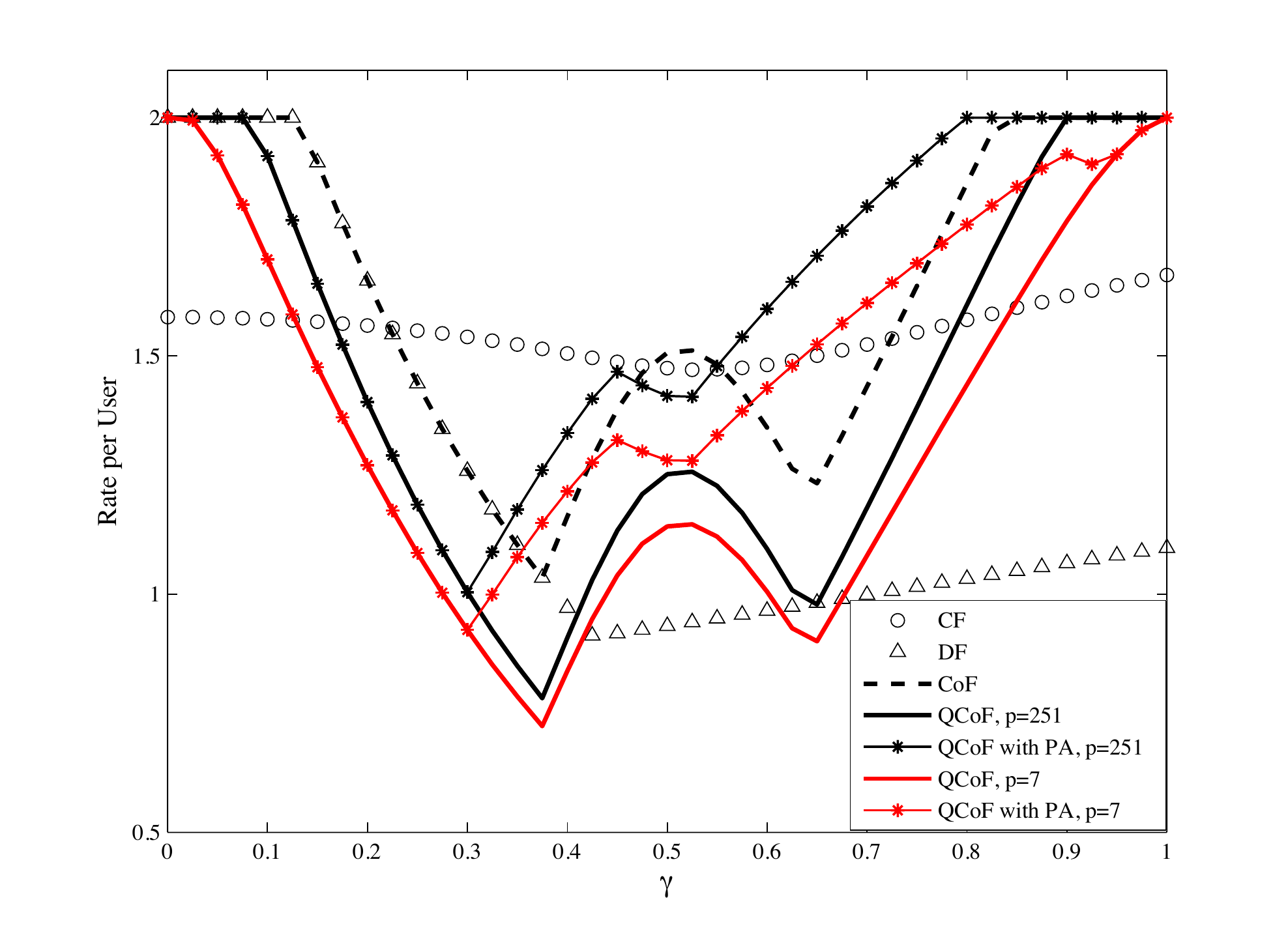}}
\caption{$\SNR=15$dB. Achievable rates per user as a function of the inter-cell interference level $\gamma$,
for finite backhaul capacity $R_{0}=2$ bits and ADC level, $p=7$ or $251$.}
\label{simulation_BC}
\end{figure}

It is known that the performance of CoF (and therefore of QCoF) is quite sensitive to the channel coefficients, due to the non-integer penalty term.
More favorable channel coefficients can be obtained by using a \emph{power allocation} (PA) strategy.
to reduce the impact of non-integer error term, which is simpler than the superposition strategy in \cite{Sanderovich2}.
Odd-numbered UTs transmit at power $\beta P$ and even-numbered UTs transmit at power $(2-\beta)P$, for $\beta \in [0,1]$.
The role of odd- and even-numbered UTs is reversed at alternate time slots, such that each UT satisfies its individual power constraint on average.
Accordingly, the effective coefficients of the channel for odd-numbered and even-numbered relays are
$\hv_{o} = [\gamma \sqrt{2-\beta}, \sqrt{\beta}, \gamma\sqrt{2-\beta}]$
and $\hv_{e}=[\gamma \sqrt{\beta}, \sqrt{2-\beta}, \gamma\sqrt{\beta}]$.
For a given $\gamma$, the parameter $\beta \in [0,1]$ can be optimized to make the effective channels better suited for the integer approximation.
In this case, the rate achieved by QCoF with PC is $R = \min\{R',R_{0}\}$ where
\begin{equation*}
R' = \max_{\beta \in [0,1]} \min \left \{ \max_{\av, \alpha} R(\hv_o, \av, \alpha) , \max_{\av,\alpha} R(\hv_e, \av,\alpha)  \right \}.
\end{equation*}
Notice that the odd- and even-numbered relays can optimize their own equation coefficients independently, but
the optimization with respect to $\beta$ is common to both and the computation rate is the minimum computation
rate over all the relays, since the same code $\Cc$ is used across all users.
In Fig.~\ref{simulation_BC}, we show the system performance for $R_{0} = 2$ bits, $p=7$ or $251$ (e.g., about $3$ or $8$ bits ADC) and
compare various relaying strategies.  The achievable rates for DF, CF, and CoF are as in \cite{Sanderovich2}, for the sake of comparison.

Not surprisingly, QCoF only pays the shaping gain with respect to CoF when $p=251$. Also, QCoF with $p=7$ shows the satisfactory performance with lower complexity. The PA strategy significantly reduces the integer approximation penalty and improves the achievable rate in the middle range of $\gamma$. Notice also that Fig.~\ref{simulation_BC} does not provide a fair comparison, since the impact of a finite resolution ADC is not considered in DF, CF, and CoF.

\end{document}